\documentclass[twocolumn, pra, showpacs,superscriptaddress]{revtex4}
\usepackage{graphicx}
\usepackage{dcolumn}
\usepackage{bm}
\usepackage{amsmath}

\renewcommand {\vec} {\mathbf}
\begin{document}

\title{Phase separation in the trapped spinor gases with anisotropic spin-spin interaction}
\author{Yajiang Hao}
\affiliation { Beijing National Laboratory for Condensed Matter
Physics, Institute of Physics, Chinese Academy of Sciences,
Beijing 100080, P. R. China}
\affiliation{Department of Physics and
Institute of Theoretical Physics, Shanxi University, Taiyuan
030006, P. R. China}
\author{Yunbo Zhang}
\affiliation{Department of Physics and Institute of Theoretical
Physics, Shanxi University, Taiyuan 030006, P. R. China}
\author{J. Q. Liang}
\affiliation{Department of Physics and Institute of Theoretical
Physics, Shanxi University, Taiyuan 030006, P. R. China}
\author{Shu Chen}
\email{schen@aphy.iphy.ac.cn}
\affiliation{ Beijing National
Laboratory for Condensed Matter Physics, Institute of Physics,
Chinese Academy of Sciences, Beijing 100080, P. R. China}
\begin{abstract}
We investigate the effect of the anisotropic spin-spin interaction
on the ground state density distribution of the one dimensional
spin-1 bosonic gases within a modified Gross-Pitaevskii theory
both in the weakly interaction regime and in the Tonks-Girardeau
(TG) regime. We find that for ferromagnetic spinor gas the phase
separation occurs even for weak anisotropy of the spin-spin
interaction, which becomes more and more obvious and the component
of $m_F=0$ diminishes as the anisotropy increases. However, no
phase separation is found for anti-ferromagnetic spinor gas in
both regimes.
\end{abstract}
 \maketitle

\preprint{APS/123-QED}

\narrowtext

Since Bose-Einstein condensates (BECs) of trapped alkali atomic
clouds were realized experimentally \cite{Anderson}, many new
regimes have been investigated extensively. The experimental
progress on trapping cold atoms under a highly controllable way has
opened the exciting opportunities for studying strongly correlated
atomic systems. When BECs are confined in a far-off-resonant optical
trap regardless of their hyperfine state, the atomic spin degrees of
freedom are liberated and the spinor nature of the condensate can be
manifested \cite {spinor1st}. It stimulates enormous theoretical and
experimental interests in studying a variety of spin-related
properties, such as quantum entanglement of spins, spin domains, etc
\cite {T.L.Ho1,LYou}. Especially, the magnetism of the spinor gas
has been studied by many authors \cite{T.L.Ho1,Han Pu,Q Gu,Recati}.
A great number of theories and experiments have shown that when the
spinor gas realized in a magnetic trap is loaded into an optical
trap, the spin domain will form after evolution for a period of
time.

Domain formation or phase separation in the multicomponent BECs
was also intensively investigated in the past years. The
condensate mixture displays a novel phase in which phase
separation occurs if there exists a strong repulsive interaction
between the species \cite{ps}. For the spinor gases, the
spin-dependent interaction is much weaker than the contact
interaction, and thus the spin-dependent interaction almost has no
effect on the total density distribution. Although each component
displays a different density profile, no phase separation was
found in the case of the isotropic spin-spin interaction\cite{WX
Zhang,self}. In this paper, we will show that anisotropic
spin-spin interaction could result in the formation of the static
spin domain in the spinor gases and we mainly focus on
one-dimensional (1D) cold atom systems for both the weakly
interacting regime and the strongly interacting TG regime.

Recently, there has been tremendous experimental progress towards
the realization of trapped 1D cold atom systems \cite
{gorlitz,Paredes,Toshiya,esslinger}. An array of 1D quantum gas is
obtained by tightly confining the particle motion in two
directions to zero point oscillations \cite{Ketterler} realized by
means of two-dimensional optical lattice potentials. By loading
BECs in the optical lattice or changing the trap intensities, and
hence the atomic interaction strength, the atoms can be made to
act either like a condensate or like a TG gas
\cite{Girardeau,Chen2}. The important parameter characterizing the
different physical regimes of the 1D quantum gas is $\gamma
=mg/\hbar ^2\rho $, the ratio of the interaction to kinetic
energy, where $g$ is an effective 1D interaction constant, $m$ is
the mass of the atom, and $\rho $ is the density.

Let us consider a repulsively interacting spin-1 Bose condensate
trapped by a harmonic potential that does not depend on the atomic
internal states
$V(\vec{r})=\frac{m}{2}[\omega_{x}^2x^2+\omega_{\perp}^2(y^2+z^2)]$,
where $m$ is the mass of each boson, $\omega_{x}$ is the trapping
frequency along the $x$ (radial) direction, and
$\omega_{y}=\omega_{z} \equiv \omega_{\perp}$ is the trapping
frequency along the $y$ and $z$ (transverse) directions. Assuming
the radial confinement ($\hbar \omega_{x}$) much weaker than the
transversal one ($\hbar \omega_{\perp}$) leads to a $1$D
configuration, in which the motion of the atoms is frozen along
the transverse directions. In such a situation, the external
potential that contributes to the atomic motion reads
$V(x)=\frac{m}{2}\omega_{x}^2x^2$. In second quantization
language, the Hamiltonian of our system may be expressed as
\begin{equation}
\mathcal{H}=H_0+H_{int}+H_{spin}  \label{Hspinor}
\end{equation}
with
\begin{eqnarray*}
H_0 &=&\int dx\hat{\Psi}_i^{\dag}\left( -\frac{\hbar ^2}{2m}\frac{d^2}{dx^2}%
+V\left( x\right) \right) \hat{\Psi}_i, \\
H_{int} &=&\frac{c_0}2\int dx:\left(
\hat{\Psi}_i^{\dag}\hat{\Psi}_i\right) ^2:,
\end{eqnarray*}
where we are assuming that the interaction between bosons is
described by a contact two-body potential, which may be described by
a Dirac-delta function. Finally,
\begin{eqnarray*}
H_{spin} &=&\frac{c_2}2\int
dx\hat{\Psi}_k^{\dag}\hat{\Psi}_i^{\dag}[\left(
F_z\right) _{ij}\left( F_z\right) _{kl}+ \\
&&\Delta \left( \left( F_x\right) _{ij}\left( F_x\right)
_{kl}+\left( F_y\right) _{ij}\left( F_y\right) _{kl}\right)
]\hat{\Psi}_j\hat{\Psi}_l
\end{eqnarray*}
describes the spin-spin interaction. In $H_{spin}$, the anisotropy
parameter $\Delta$ is introduced phenomenologically to describe the
anisotropy of spin-spin interaction, which may arise from the
anisotropic magnetic dipole-dipole interaction, whereas $\Delta=1$
corresponds to the isotropic model. When $c_2<0$ the system is in a
ferromagnetic regime, while for $c_2>0$, the system is
anti-ferromagnetic. We limit our discussion on $0\leq \Delta \leq 1$
because only in this regime the phase separation may take place
\cite{remark}. Here $\hat{\Psi}_i\left( x\right) $
($\hat{\Psi}_i^{\dag }\left( x\right) $ ) is the field operator that
annihilates (creates) an atom in the $i$-th internal state at
location $ x$, $i=+,0,-$ denotes the atomic hyperfine state $\left|
F=1,m_F=+1,0,-1\right\rangle $, respectively. Summation is assumed
for repeated indices in the above Hamiltonian and the pair of colons
denote the normal-order product. $F_x$, $F_y$ and $F_z$ are the
spin-1 matrices with the quantization axis taken along the $z$-axis
direction. The atomic interaction constants are expressed through
the effective 1D interaction strength $U_{0,2}$ with $
c_0=\frac{U_0+2U_2}3 $ and $ c_2=\frac{U_2-U_0}3$, where $U_{0,2}$
have the following relation
\begin{eqnarray}
U_{0,2} &=&-\frac{2\hbar ^2}{ma_{0,2}^{1D}}, \\
a_{0,2}^{1D} &=&-\frac{d_{\bot }^2}{2a_{0,2}}\left(
1-\mathcal{C}\left( a_{0,2}/d_{\bot }\right) \right).   \nonumber
\end{eqnarray}
Here  $a_{0,2}$ denotes the $s$-wave scattering lengths between two
identical spin-1 bosons in the combined symmetric channel of total
spin $0(2)$ when the cold atoms are trapped intensively in
transverse direction with the transverse trapping frequency $\hbar
\omega _{\bot }$ \cite{self,Olshanii,Olshanii2,Petrov}, $d_{\bot
}=\sqrt{\hbar /m\omega _{\bot }}$ and $\mathcal{C}\approx 1.4603$.

In order to deal with the weakly and strongly interacting regimes
on the same footing, we work in a scheme of modified
Gross-Pitaevskii theory \cite{Dunjko,self,Chen,Kolomeisky} in
which the energy density $\epsilon \left( \rho \right) $ is taken
from the exactly solvable problem of a three-component Bose
gas\cite{self}. It follows that the properties of a spinor gas are
determined by the following spin-dependent energy functional
\begin{eqnarray}
\mathcal{E} &=&\int dx\left[ \Phi _i^{*}\left( -\frac{\hbar
^2}{2m}\frac{d^2 }{dx^2}+V\left( x\right) \right) \Phi _i+\rho
\epsilon \left( \rho
\right) \right]   \nonumber \\
&&+\int dx\frac{c_2}2\Phi _k^{*}\Phi _i^{*}[\left( F_z\right) _{ij}\left(
F_z\right) _{kl}+  \label{EF} \\
&&\Delta \left( \left( F_x\right) _{ij}\left( F_x\right) _{kl}+\left(
F_y\right) _{ij}\left( F_y\right) _{kl}\right) ]\Phi _j\Phi _l,  \nonumber
\end{eqnarray}
where $\rho =\sum_i\rho _i=\sum_i\left| \Phi _i\right| ^2$ and the
energy density\cite{Lieb,Chen,Sutherland}
\begin{equation}
\epsilon \left( \rho \right) =\frac{\hbar ^2}{2m}\rho e\left( \gamma \right)
=\{
\begin{array}{ll}
\text{ \ \qquad }c_0\rho /2, & \gamma \ll 1 \\
\pi ^2\hbar ^2\rho ^2/6m, & \gamma \gg 1
\end{array}
.
\end{equation}
The first line of energy functional (3) is made up of three
contributions: the first one derives from the usual kinetic energy
operator, the second one represents the additional term related to
the inhomogeneity due to the external confinement $V(x)$
\cite{Lieb2003}, and the last one corresponds to the energy density
in the homogeneous system. In the second and third lines, the
contribution deriving from spin-spin interaction is involved; in
particular, the term in the square brackets has the explicit form
\begin{eqnarray}
&&\Delta \left( 2\rho _0\rho _{-}+2\rho _{+}\rho _0+2\Phi _0^{*2}\Phi
_{+}\Phi _{-}+2\Phi _0^2\Phi _{-}^{*}\Phi _{+}^{*}\right)   \nonumber \\
&&+\rho _{+}^2+\rho _{-}^2-2\rho _{+}\rho _{-}.
\end{eqnarray}
At this point, we stress that, generally, the spin-spin interaction
coupling $(c_2)$ constant is much smaller than the $s$-wave
interaction one $(c_0)$, i.e. $c_2<<c_0$. If we assume that there is
no spin-spin interaction $(c_2=0)$ and no external trapping
$(V(x)=0)$, then the system described by Hamiltonian (1) is
integrable \cite {Sutherland}; its ground-state energy density has
the same form as that of Lieb-Liniger problem \cite{Lieb}.

In the system both the total atom number and the magnetization
$\mathcal{M} =\int dx\left\langle F\right\rangle =\int dx\left[ \Phi
_{+}^{*}\Phi _{+}-\Phi _{-}^{*}\Phi _{-}\right] $ are conserved
\cite{S. Yi,WX Zhang}. In order to obtain the ground state from a
global minimization of $\mathcal{E}$ with the constraints on both
$N$ and $\mathcal{M}$, we introduce separately Lagrange multiplier
$B$ to conserve $M$ and the chemical potential $\mu $ to conserve
$N$. The ground state is then determined by a minimization of the
free-energy functional $\mathcal{F}=\mathcal{E}-\mu N-B\mathcal{M}$.
The dynamics of $\Phi _i$ is governed by the coupled GPEs
\begin{eqnarray}
i\hbar \frac{\partial \Phi _{+}}{\partial t} &=&\left[ H-B+c_2\left( \rho
_{+}+\Delta \rho _0-\rho _{-}\right) \right] \Phi _{+}+c_2\Delta \Phi
_0^2\Phi _{-}^{*},  \nonumber \\
i\hbar \frac{\partial \Phi _0}{\partial t} &=&\left[ H+c_2\left( \rho
_{+}+\rho _{-}\right) \right] \Phi _0+2c_2\Delta \Phi _{+}\Phi _{-}\Phi
_0^{*},  \label{GPEs} \\
i\hbar \frac{\partial \Phi _{-}}{\partial t} &=&\left[ H+B+c_2\left( \rho
_{-}+\Delta \rho _0-\rho _{+}\right) \right] \Phi _{-}+c_2\Delta \Phi
_0^2\Phi _{+}^{*},  \nonumber
\end{eqnarray}
with
\begin{equation}
H=-\frac{\hbar ^2}{2m}\frac{d^2}{dx^2}+V\left( x\right) +\tilde{F}%
\left( \rho \right)
\end{equation}
and
\begin{equation}
\tilde{F}\left( \rho \right) =\frac \partial {\partial \rho }\left[ \rho
\epsilon \left( \rho \right) \right] =\{
\begin{array}{ll}
\text{ \ \qquad }c_0\rho , & \gamma \ll 1 \\
\pi ^2\hbar ^2\rho ^2/2m, & \gamma \gg 1
\end{array}
.
\end{equation}
We obtain the ground state of spin-$1$ BECs by propagating the
coupled GPEs Eq. (\ref{GPEs}) in imaginary time. In each propagating
step, the wave function $\Phi _i$ is normalized to conserve the
atomic number and by adjusting the Lagrange multiplier $B$ the
conservation of magnetization $ \mathcal{M}$ is assured. In our
procedure the Crank-Nicholson scheme is used. We will determine the
ground state for the 1D spinor Bose gases trapped in the harmonic
trap $V\left( x\right) =\frac 12m\omega_x ^2x^2$ both in the weakly
interacting regime and in the TG regime.
\begin{figure}[tbp]
\includegraphics[width=3.5in]{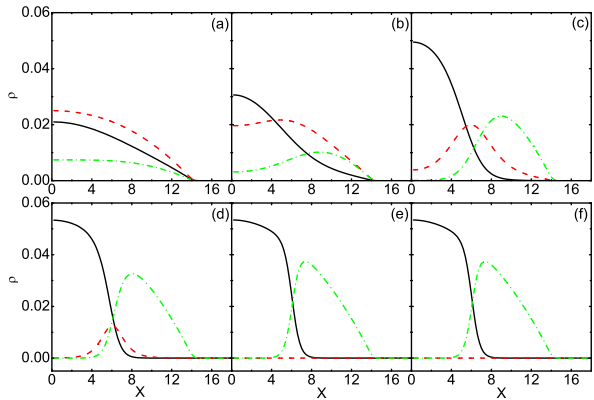}
\caption{The density profile of ground state of a spin-1 $^{87}$Rb
condensates for the + (solid line) component, 0 (dashed line)
component and - (dash-dot lines) component in the weakly interaction
regime with m=0.2. (a) $\Delta$=1.0; (b) $\Delta$=0.9; (c)
$\Delta$=0.8; (d) $\Delta$=0.5; (e) $\Delta$=0.2; (f) $\Delta$=0.0.
In this figure the length is in the unit of $a=1.2\mu$m.}
\label{fig1}
\end{figure}
\begin{figure}[tbp]
\includegraphics[width=3.5in]{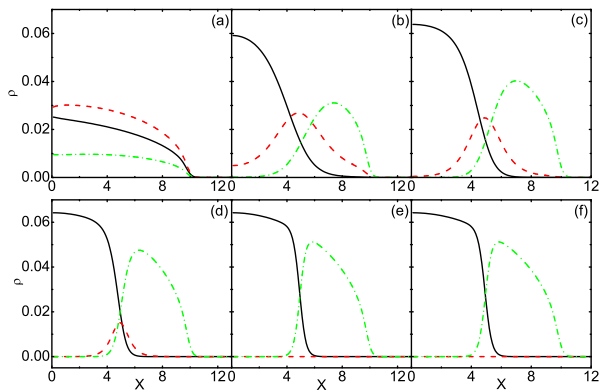}
\caption{The density profile of ground state of a spin-1 $^{87}$Rb
condensates for the + (solid line) component, 0 (dashed line)
component and - (dash-dot lines) component in the Tonks regime with
m=0.2. (a) $\Delta$=1.0; (b) $\Delta$=0.9; (c) $\Delta$=0.8; (d)
$\Delta$=0.5; (e) $\Delta$=0.2; (f) $ \Delta$=0.0. In this figure
the length is in the unit of $a=8.5\mu$m.} \label{fig2}
\end{figure}

To investigate the effect of of anisotropy parameter $\Delta $, we
evaluate the density profile of the ground state of 1D spinor gases
for the $^{87}$Rb (ferromagnetic) with $a_0=102a_B$ and $a_2=100a_B$
($a_B$ is the Bohr radius) in the harmonic trap for different
anisotropy parameters. By properly tuning the parameters, the system
may be either in the weakly interacting regime or in the TG regime.
Let us first consider the specific system with the typical
parameters of the trap $\omega _x=0.5$kHz, $\omega _{\bot }=50$kHz
and the atomic number $N=2000$, in which case the effective
interaction strength $ \gamma \sim 0.008$ indicating that the system
is in the weakly interacting regime. Fig. 1 displays the density
profiles in unites of $N/a$ with $a=\sqrt{\hbar /{m\omega_x} }$ for
different anisotropy parameter $\Delta $ in the weakly interaction
regime with m$= \frac{\mathcal{M}}N=0.2$. When the spin-spin
interaction is isotropic, i.e. $ \Delta =1.0$, the three different
components superpose each other and they have similar distributions.
In the case of weak anisotropy, for instance $\Delta =0.9$ here, the
distribution has changed explicitly. Therefore the ground state
configurations have positive magnetization in some region but
negative in another. Also the $0$ component (dashed line) diminishes
as the anisotropy increases. As the anisotropy becomes more and more
clear, the components tend to separate and coincide only at the
boundary between them. The components always try to avoid each
other. It is shown that when $\Delta =0.2$ the $0$ component
disappears completely and the density profiles of $+$ component
(solid line) and $-$ component (dash-dot lines) exhibit in the form
of phase separation. The corresponding density profiles in the TG
regime are plotted in Fig. 2 with m$=0.2$. In this regime the
parameters are tuned to $\omega _x=10$ Hz, $\omega _{\bot }=500$kHz,
the atomic number $N=50$ and the effective interaction strength
$\gamma \sim 15$. In this case, the similar density distributions
occur. Comparing the Fig. 1(a) and the Fig.2(a) we see that the
density profiles in the TG regime behave like that of Fermions.
According to Fig. 2(b), with the very weak anisotropy, obvious phase
separation has occurred in the Tonks regime. Fig. 3 displays the
density profiles for different magnetization but with the same
anisotropy parameter, which indicates that the magnetization only
influences on the ratio of the atomic numbers between the $+$
component and $-$ component. It turns out that although the
spin-spin interaction is much small, its property affects greatly
the density distribution of each component of the spinor gas and the
phase separation occurs more easily in the TG regime than in the
weakly interacting regime. The 1D spinor gas in the TG regime might
provide us a good platform to investigate the magnetism of the cold
atoms.
\begin{figure}[tbp]
\includegraphics[width=3.7in]{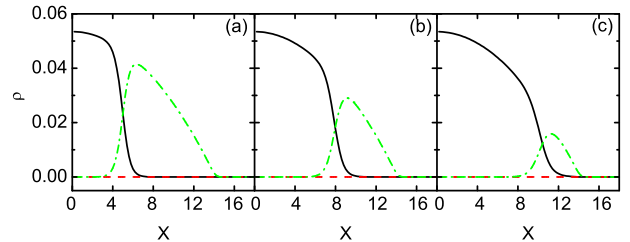}
\caption{The density profile of ground state of a spin-1 $^{87}$Rb
condensates for the + (solid line) component, 0 (dashed line)
component and - (dash-dot lines) component in the weakly interaction
regime with $\Delta $=0.2. (a) m=0.0; (b) m=0.5; (c) m=0.8. In this
figure the length is in the unit of $a=1.2\mu$m.} \label{fig3}
\end{figure}

For further studying the effect of anisotropy on the
anti-ferromagnetic system, we consider a condensate of $^{23}$Na
in the weakly interacting regime with$\ a_0=50a_B$ and
$a_2=55.1a_B$. The trap frequencies are chosen as $\omega_x=50$Hz,
$\omega _{\bot }=10$kHz and $N=1000$ so that $\gamma \sim 0.001$.
The density profiles are shown in Fig. 4 with $m=0.2$ for
anisotropic and isotropic case. It is shown that no phase
separation occurs even for very large anisotropy and the same
distribution as the case of isotropic spin-spin interaction
displays \cite{self}. From the above results, it is obvious that
the density distributions of every component strongly depends on
its' anisotropy and the ferromagnetic or the anti-ferromagnetic
properties of the relatively weak spin interactions, whereas the
total density is almost not affected by the weak spin
interactions.
\begin{figure}[tbp]
\includegraphics[width=3.5in]{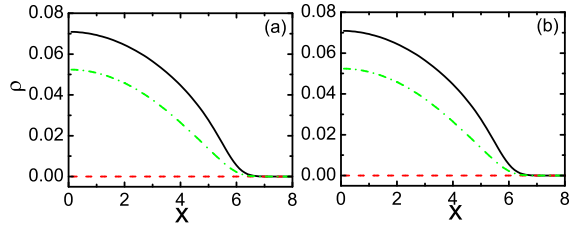}
\caption{The density profile of ground state of a spin-1 $^{23}$Na
condensates for the + (solid line) component, 0 (dashed line)
component and - (dash-dot lines) component in the weakly interaction
regime with m=0.2. (a) $\Delta $=0.5; (b) $\Delta $=1.0. In this
figure the length is in the unit of $a=7.4\mu$m.} \label{fig4}
\end{figure}

Finally, we discuss the possible experimental realization of
anisotropic spin interaction in spinor gases. It is well known
that magnetic dipolar interactions are anisotropic despite the
fact that dipolar interactions are rather weak comparing to the
spin interactions. The relative strength of the dipolar and the
spin exchange interactions is estimated to be $10^{-1}$ for
$^{87}$Rb and $10^{-3}$ for $^{23}$Na \cite{S. Yi2}. Since the
magnetic dipole-dipole interaction is irrelevant to the $s$-wave
scattering length,  we may tune the $s$-wave scattering length
experimentally by the Feshbach resonance so that the strength of
isotropic spin-spin interaction $c_2$ is comparable to that of
dipole-dipole interaction, and thus the anisotropic interaction
becomes obvious. Although a condensate of strongly anisotropic
spinor gases remains to be realized, experimentalists in
Ref.\cite{L. Santos} have already successfully demonstrated the
effect of the dipole-dipole interaction. With the present rapid
development in the experimental manipulation of cold atoms, the
goal of making a condensate of spinor gases with anisotropic spin
interaction does not seem to be far-fetched. It is worth
indicating that our approach can be directly applied to deal with
the three-dimensional (3D) problem for which the mean-field theory
corresponds to our weakly interacting theory. It follows that a 3D
spinor gas with anisotropic spin interactions also displays phase
separation. However, the result for the strongly interacting
regime can not be extended to the higher-dimensional case in which
no TG gas can be realized. Our work is helpful to understand the
properties of the spinor condensates and deepen our understanding
of formation of spin domains in spinor gases.

In summary, we have studied the density profile of the ground
states of 1D spin-$1$ Bose gases for different anisotropy
parameter $\Delta $. The distributions of the ferromagnetic spinor
gas are affected tremendously by $\Delta $ although the $c_2$ term
is very small compared with $c_0$ term in Eq. (\ref{Hspinor}).
Even if the anisotropy is weak, the distributions show obvious
difference from that of the isotropic case. In the large
anisotropy the component of $m_F=0$ disappears and obvious phase
separation occurs both in the weakly interaction regime and in the
Tonks regime. And the effect of anisotropy in the TG regime can
display more obviously with weaker anisotropy than the former
case. However, when the spinor gas is anti-ferromagnetic, the
distribution is no longer being affected by the anisotropy
parameter.

This work was supported by the NSF of China under Grant No.
90203007 and Grant No. 10574150. We thank S. Yi, L. You, Y. Wang,
W.-X. Zhang and W.-D. Li for useful discussions.


\begin{thebibliography}{99}
\bibitem{remark} We note that $\Delta=1$ corresponds to the case with no
anisotropy and, in principle, $|\Delta|>1$ is also possible. We
investigated the case with $|\Delta|>1$ but found that no phase
separation occurs.
\end{thebibliography}
\end{document}